\documentclass{aa}  

\usepackage{graphicx}
\usepackage{txfonts}
\bibpunct{(}{)}{;}{a}{}{,} 

\begin{document}

   \title{Infrared-faint radio sources are at high redshifts}

   \subtitle{Spectroscopic redshift determination of infrared-faint radio
   sources using the \textit{Very Large Telescope}}

   \author{A. Herzog\inst{1,2,3}
          \and
          E. Middelberg\inst{1}
	  \and
	  R. P. Norris\inst{3,4}
	  	  \and
	  R. Sharp\inst{5}
	  \and
	  L. R. Spitler\inst{2,6}
	  \and
	  Q. A. Parker\inst{2,6}
          }

   \institute{Astronomisches Institut, Ruhr-Universit\"at Bochum, Universit\"atsstr. 150, 44801 Bochum, Germany\\
              \email{herzog@astro.rub.de}
         \and
             Macquarie University, Sydney, NSW 2109, Australia
         \and
            CSIRO Astronomy and Space Science, Marsfield, PO Box 76, Epping, NSW
            1710, Australia
         \and
            ARC Centre of Excellence for All-sky Astrophysics (CAASTRO)
         \and
         	Research School of Astronomy \& Astrophysics, Australian National University, Mount Stromlo Observatory, Cotter road, Weston Creek, ACT 2611, Australia
         \and
            Australian Astronomical Observatory, PO Box 915, North Ryde, NSW
            1670, Australia
  }

   \date{Received 30 November 2013; accepted 16 June 2014}

 
  \abstract
   {Infrared-faint radio sources~(IFRS) are characterised by relatively high
   radio flux densities and associated faint or even absent infrared and
   optical counterparts. The resulting extremely high radio-to-infrared flux density
   ratios up to several thousands were previously known only for high-redshift
   radio galaxies~(HzRGs), suggesting a link between the two classes of object.
   However, the optical and infrared faintness of IFRS makes their study
   difficult. Prior to this work, no redshift was known for
   any IFRS in the Australia Telescope Large Area Survey~(ATLAS) fields
   which would help to put IFRS in the context of other classes of object,
   especially of HzRGs.
   }
   {This work aims at measuring the first redshifts of IFRS in the
   ATLAS fields. Furthermore, we test the hypothesis that IFRS are similar to
   HzRGs, that they are higher-redshift or dust-obscured versions of these
   massive galaxies.}
   {A sample of IFRS was spectroscopically observed using the Focal Reducer and
   Low Dispersion Spectrograph~2~(FORS2) at the Very Large Telescope~(VLT).
   The data were calibrated based on the Image Reduction and Analysis
   Facility~(IRAF) and redshifts extracted from the final spectra, where
   possible. This information was then used to calculate rest-frame luminosities,
   and to perform the first spectral energy distribution modelling of IFRS
   based on redshifts.
   }
   {We found redshifts of 1.84, 2.13, and 2.76, for three IFRS,
   confirming the suggested high-redshift character of this class of object.
   These redshifts and the resulting luminosities show IFRS to be similar
   to HzRGs, supporting our hypothesis. We found further evidence
   that fainter IFRS are at even higher redshifts.}
   {Considering the similarities between IFRS and HzRGs substantiated in
   this work, the detection of IFRS, which have a significantly higher sky density
   than HzRGs, increases the number of active galactic nuclei in the early
   universe and adds to the problems of explaining the formation of supermassive
   black holes shortly after the Big Bang.}


   \maketitle
%

\section{Introduction}
\label{introduction}
Active galactic nuclei~(AGN) at high redshifts are an important field
of current research since they play a crucial role in answering
basic questions about the evolution of the universe. For example,
high-redshift AGN are used to study the evolution of the link between
black hole mass and the properties of its host galaxy~(e.g.
\citealp{Hopkins2005}, \citealp{Lamastra2010}), their impact on the
reionisation and the structure formation in the universe~(e.g.
\citealp{Fan2006}, \citealp{Robertson2010}, \citealp{Boutsia2011}), and
the growth of supermassive black holes~(SMBHs) with masses $>10^9
M_\odot$ in less than one billion years after the Big Bang~(e.g.
\citealp{VolonteriRees2005}). Infrared-faint radio sources -- a class
of object whose detection was unexpected -- could
significantly contribute to the population of high-redshift AGN.\par

\subsection{Definition and discovery of IFRS}
\label{definition}

Infrared-faint radio sources~(IFRS) are peculiar objects, characterised by
relatively high radio flux densities on the order of 1\,mJy at 1.4\,GHz
only associated with faint or even absent infrared~(IR) counterparts. They are
defined by \citet{Zinn2011} by two criteria:

\begin{enumerate}[(i)]
  \item radio-to-IR flux density ratio
    $S_{1.4~\textrm{GHz}}/S_{3.6~\mu\textrm{m}} > 500$ and

  \item 3.6~$\mu$m flux density $S_{3.6~\mu\textrm{m}} < 30~\mu$Jy~.
\end{enumerate}

The first criterion selects objects that are clear outliers from the
radio-IR correlation, while the second criterion prevents the selection of
low-redshift objects with $z\lesssim 1.4$~\citep{Zinn2011}.\par

The discovery of IFRS was unexpected. When \citet{Norris2006} crossmatched the
deep 1.4~GHz radio maps from the Australia Telescope Large Area Survey~(ATLAS)
with the data from the \textit{Spitzer} Wide-area Infrared Extragalactic
Survey~(SWIRE; \citealp{Lonsdale2003}), it was expected that SWIRE would provide
an IR counterpart for any extragalactic radio source detected in ATLAS. The
spectral energy distribution~(SED) of typical host galaxies at redshifts
below~4, regardless of whether the radio emission is dominated by AGN or star
forming activity, should place the IR flux above the SWIRE detection
limit~\citep{Norris2006}. However, 22~sources were found in the Chandra Deep
Field South~(CDFS) without an IR counterpart in the crossmatching process. These
sources -- constrained at 3.6\,$\mu$m by a $5\sigma$ upper limit of $5\,\mu$Jy
-- were labelled as IFRS. Using the same approach,
\citet{Middelberg2008ELAIS-S1} identified 31~IFRS using ATLAS radio observations
of the European Large Area IR space observatory Survey South~1~(ELAIS-S1) field
and SWIRE data. In total, 53~IFRS were detected in the overlapping region
between SWIRE and ATLAS, with 1.4~GHz flux densities between tenths and tens of
mJy. \citet{Zinn2011} noted that none of those IFRS has an X-ray counterpart in
the XMM-Newton survey of the ELAIS-S1 field~\citep{Puccetti2006}. One IFRS is
located in the field of the CDFS \textit{Chandra}~2\,Ms
survey~(\citealp{Giacconi2002,Luo2008}; minimum full band sensitivity around
$3.3\times 10^{-16}$\,erg\,s$^{-1}$\,cm$^{-2}$ over the Great Observatories
Origins Deep Survey-South field). However, this IFRS also remains undetected in
the X-ray regime.\par

\subsection{VLBI observations and SED modelling}

The first attempts to further characterise the properties of IFRS used
Very Long Baseline Interferometry~(VLBI)
observations. \citet{Norris2007} observed two IFRS and identified an
AGN in one of them, showing a linear size of $\leq 260$~pc at any
redshift. Furthermore, \citet{Middelberg2008IFRS_VLBI} detected one of
four targeted IFRS with VLBI and derived a brightness temperature of
$3.6\times 10^6$~K, implying non-thermal emission from an AGN since
thermal emission processes cannot produce such high brightness
temperatures. Therefore, \citeauthor{Norris2007} and
\citeauthor{Middelberg2008IFRS_VLBI} concluded that at least a
fraction of all IFRS contain AGN.\par

\citet{GarnAlexander2008} found a sample of 14~IFRS in the
\textit{Spitzer} First Look Survey~(FLS) field, using \textit{Spitzer}
Infrared Array Camera~(IRAC; \citealp{Fazio2004}) and Multiband
Imaging Photometer~(MIPS; \citealp{Rieke2004}) data. They concluded
from SED modelling that IFRS are probably 3C~sources redshifted to
$2\leq z\leq 5$. Moreover, they excluded obscured star forming
galaxies~(SFGs) as an explanation for the objects in their sample
because of the radio-to-IR flux density ratio upper limits, differing
significantly from corresponding values of SFGs.\par

The first SED modelling of ATLAS-IFRS was presented by \citet{Huynh2010}, using
new ultra-deep imaging in the extended CDFS~(eCDFS). From their detailed SED
modelling of four IFRS, \citeauthor{Huynh2010} concluded that a 3C\,273-like
object can reproduce the data when redshifted to $z>2$.  In agreement with
\citet{GarnAlexander2008}, \citeauthor{Huynh2010} showed that all four analysed
IFRS fall well beyond the radio-IR correlation, suggesting the radio emission is
produced by the presence of an AGN, but not by star forming activity.\par

\subsection{Similarities between IFRS and HzRGs}

\citet{Middelberg2011} studied the radio properties of 17 out of all 31~IFRS
from the ATLAS ELAIS-S1 sample between 2.3\,GHz and 8.4~GHz. They found a median
radio spectral index of $\alpha=-1.4$\footnote{The spectral index is defined as
$S\propto\nu^\alpha$.} and no index larger than $-0.7$, which is significantly
steeper than the radio spectra of the general source
population~$(-0.86)$ and of the AGN source
population~$(-0.82)$ in the ATLAS ELAIS-S1 field. Furthermore,
\citeauthor{Middelberg2011} noticed similarities between IFRS and the sample of
high-redshift radio galaxies~(HzRGs) from \citet{Seymour2007}. These HzRGs show
steep radio spectra (median radio spectral index of $-1.02$) like IFRS.
Moreover, the extremely high radio-to-IR flux density ratios of IFRS overlap
with those of this HzRG sample. Recently, \citet{Singh2014arXiv} searched for
HzRGs in faint ultra steep spectrum~(USS) radio sources and found several
sources overlapping with the selection criteria of IFRS.\par

The hypothesis of an AGN existing in at least a fraction of all IFRS was
supported by \citeauthor{Middelberg2011} who identified ten~IFRS of their sample
as AGN and the other seven as probable AGN, based on radio-to-IR flux density
rations, polarisation properties, radio spectral indices, VLBI detections, or
radio morphology.\par

These similarities between IFRS and HzRGs were emphasised by \citet{Norris2011}
who showed that no other type of object occupies the range of radio-to-IR flux
density ratios of IFRS except for HzRGs. The sample of HzRGs of
\citet{Seymour2007} shows a relation between the $3.6~\mu$m flux density and
redshift similar to the $K-z$ relation for other radio
galaxies~\citep{Willott2003}.
\citeauthor{Norris2011} suggested that IFRS might follow the same correlation.
They used deep data from the \textit{Spitzer} Extragalactic Representative
Volume Survey~(SERVS; \citealp{Mauduit2012}) with a $3\sigma$ noise level of
$\sim 1.5~\mu$Jy at $3.6~\mu$m in the fields of CDFS and ELAIS-S1 and stacked
the $3.6~\mu$m images of 39~IFRS, which resulted in an upper flux density limit
of $\sim 0.2~\mu$Jy, but remained without a detection. This was interpreted as
evidence for the significant IR faintness of IFRS.\par

\subsection{IFRS catalogues}

\citet{Zinn2011} compiled a catalogue of 55~IFRS in four deep radio fields,
CDFS, ELAIS-S1, FLS, and Cosmological Evolution Survey~(COSMOS,
\citealp{Schinnerer2007}), based on their criteria mentioned in
Sect.~\ref{definition} and derived a survey-independent IFRS sky density of
$(30.8\pm 15.0)~\textrm{deg}^{-2}$.\par

Recently, Maini et~al. (in prep.) found 21~IFRS in the Lockman Hole and looked
for IR counterparts of IFRS located in the SERVS deep fields. 
\citet{Collier2014} used data from the all-sky survey Wide-Field Infrared Survey
Explorer~(WISE; \citealp{Wright2010}) and the Unified Radio Catalog~(URC;
\citealp{Kimball2008}) and presented a catalogue of 1317~IFRS that fulfill the
selection criteria from \citet{Zinn2011} given in Sect.~\ref{definition} but
are, on average, brighter than the IFRS found in the ATLAS fields.
\citeauthor{Collier2014} suggested that their IFRS are closer versions of the
IFRS found by \citet{Norris2006} and \citet{Middelberg2008ELAIS-S1}.\par

We note that IFRS were originally selected as radio sources without any IR
counterpart (i.e.
\citealp{Norris2006,Middelberg2008ELAIS-S1,GarnAlexander2008}). Since this
definition was survey-specific, \citet{Zinn2011} generalised this criterion to
the two survey-independent selection criteria given in Sect.~\ref{definition}.
However, these two types of selection criteria are very similar in the sense
that they select the same class of object. The definition by
\citeauthor{Norris2006} allows faint radio sources that are slightly above the
noise level and show no counterpart at 3.6\,$\mu$m to be considered as an IFRS,
whereas this is prevented by the \citeauthor{Zinn2011} criteria. On the other
hand, \citeauthor{Zinn2011} consider objects with a faint near-IR counterpart as
an IFRS, although only under the condition that the source is sufficiently
radio-bright to fulfill the radio-to-IR flux density ratio criterion.

\subsection{Current status of research}

All observational findings so far are compatible with most, if not all,
IFRS being high-redshift~($z>2$) radio-loud AGN, potentially suffering from
heavy dust extinction. Considering their high sky density, IFRS could be a very
numerous and so far overlooked population of high-redshift AGN that
could have a significant impact on the evolution of the universe as suggested
by~\citet{Zinn2011}.  Furthermore, \citeauthor{Zinn2011} concluded that the
X-ray emission of AGN-driven IFRS is consistent with the unresolved components
of the Cosmic X-ray background reported by \citet{Moretti2003}.\par

The current knowledge about IFRS is wholly based on photometric detections or
upper limits, apart from the VLBI observations. In this paper, we present the first
spectroscopic data of IFRS in the ATLAS fields. Based on these observations made
using the Focal Reducer and low dispersion Spectrograph~2~(FORS2;
\citealp{Appenzeller1998}) on the Very Large Telescope~(VLT), we present the
first redshifts of ATLAS-IFRS. Using these results we test the hypothesis that
IFRS are similar to HzRGs, derive their intrinsic properties, compare them to
other objects and model the SEDs of IFRS. The sample of observed IFRS and the
VLT FORS2 observations are described in Sect.~\ref{sample_and_observations}. In
Sect.~\ref{data_reduction_and_calibration}, we summarise the data reduction and
calibration and show the final spectra. In
Sect.~\ref{redshifts_and_intrinsic_properties_of_IFRS}, redshifts are measured
and intrinsic properties derived for our sample of IFRS. We test our hypotheses
in Sect.~\ref{discussion} and present our conclusions in
Sect.~\ref{conclusions}. The cosmological parameters used in this paper are
$\Omega_\Lambda = 0.7$, $\Omega_\textrm{M} = 0.3$, $H_0 =
70$~km~s$^{-1}$~Mpc$^{-1}$ in combination with the calculator by
\citet{Wright2006}.


\section{Sample and observations}

\label{sample_and_observations}

The sample of IFRS consists of four~objects, chosen by their detected optical
counterparts between 22.0 and 24.1~(Vega) magnitudes in the $R$~band. The
condition of an optical counterpart ensures that the sources are bright enough
for optical spectroscopy. All selected sources show a faint IR counterpart at
3.6\,$\mu$m between 14.2\,$\mu$Jy and 29.3\,$\mu$Jy, fulfilling the selection
criteria of IFRS by \citet{Zinn2011}. The presence of an IR counterpart biases
the observed sample towards less extreme IFRS, i.e. with lower radio-to-IR flux
density ratios. In the case of the four observed IFRS, this ratio is between 600
and 1100, whilst for the majority of ATLAS-IFRS, which do not show a detected IR
counterpart, lower limits of this ratio can be as high as 8000. These
IR-undetected IFRS do not show an optical counterpart and, therefore, are not
suitable for optical spectroscopy. Although the observed sample is not
representative of all IFRS, it allows us to gather information about the
intrinsic properties of less extreme IFRS and to test the hypotheses about their
relation to other classes of object. Moreover, these findings will also allow
conclusions on the more extreme IFRS which cannot be achieved by spectroscopy
because of their deep optical and IR faintness. We summarise the observed
objects and their characteristic properties in
Table~\ref{table:sample_IFRS}.\par

The longslit spectroscopic observations of these four IFRS were carried out in
project 087.B-0813(A) between July and September~2011 (ESO Period~87) in service
mode, using FORS2 on UT1 at the VLT. Using the GRISM\_150I grism, covering the
wavelength range between 330\,nm and 1100\,nm, and a slit width of 2\,\arcsec, a
high throughput is achieved at the cost of relatively low resolution, although
sufficient to provide reliable redshifts. The dispersion was 3.45~\AA/pixel,
resulting in a resolution of 54.76~\AA. The total on-source times were 44~min
for S212, 88~min for S265, 45~min for S539, and 128~min for S713, where the
exposure time for each object was split into several shorter exposures, enabling
the correction of cosmic ray events. The seeing varied during and between the
different observations from 0.86\,\arcsec to 2.46\,\arcsec.

\begin{table*}
\caption{Sample of spectroscopically observed IFRS. Listed are
  the ID, the position in RA and DEC, the radio flux density at
  1.4~GHz, the IR flux density at 3.6~$\mu$m, the radio-to-IR flux
  density ratio between 1.4~GHz and 3.6~$\mu$m, the optical $R$~band
  (Vega) magnitude, the on-source time, and the redshift determined in
  this work. Positions, flux densities, and magnitudes were taken
  from \cite{Norris2006}.}
\label{table:sample_IFRS}      
\centering                          
\begin{tabular}{c c c c c c c c c}        
\hline\hline                 
 IFRS & RA & DEC & $S_{1.4\mathrm{~GHz}}$ & $S_{\mathrm{3.6~\mu m}}$ &
 $S_{\mathrm{1.4~GHz}}/S_{\mathrm{3.6~\mu m}}$ & $R$ (Vega) & $t_\mathrm{obs}$ &
 $z$
 \\
 ID & \multicolumn{2}{c}{J2000.0} & [mJy] & [$\mu$Jy] & & [mag] & [min] & \\
\hline                        
   S212 & 03:29:48.942 & --27:31:48.98 & 18.9 & 17.5 & 1080 & 22.0 & 44 &
   $2.76\pm0.05$ \\
   S265 & 03:30:34.661 & --28:27:06.51 & 18.6 & 29.3 & 635 & 22.3 & 88 &
   $1.84\pm 0.03$ \\
   S539 & 03:33:30.542 & --28:54:28.22 & 9.1 & 14.2 & 641 & 24.1 & 45 & -- \\
   S713 & 03:35:37.525 & --27:50:57.88 & 16.4 & 25.5 & 643 & 22.4 & 128 &
   $2.13\pm 0.03$ \\ \hline                                   
\end{tabular}
\end{table*}


\section{Data reduction and calibration}

\label{data_reduction_and_calibration}

Data reduction was carried out using the standard Image
Reduction and Analysis Facility~(IRAF; \citealp{Tody1986})
procedures. All exposures were bias-corrected and flatfielded,
using normalised masterflats based on the individual dome
flat exposures, whose faulty pixels were previously corrected
using a bad pixel mask.

We rejected some single flat field exposures where an imperfect illumination was
obvious before creating the masterflat. Cosmic rays were removed using the IRAF
task \texttt{cosmicrays} and through manual inspection. Since the objects are
not located in exactly one line in the two-dimensional spectra, i.e. the spatial
position changes as a function of wavelength, we corrected the spatial axis for
this distortion. Consequently, the wavelength calibration was carried out using
lamp exposures taken in every observing night. Since all lamp exposures were
saturated and the determination of the lines' peak positions was impossible, we
applied a block average with width~3 on the spectral axis to all exposures used
in the entire reduction procedure. At the end of the wavelength calibration, the
data cover the wavelength range between 390\,nm and 1100\,nm. Consequently, we
subtracted the sky background using the IRAF task \texttt{background}.
Finally, all individual exposures of one object were averaged to increase the
signal-to-noise ratio and one-dimensional spectra were extracted by applying a
suitable aperture at the object's position using the IRAF task \texttt{apall}.
For IFRS~S265, which was observed on 2011~July~16 and 2011~July~30, we used
only the data taken on 2011~July~30 for the final spectrum because of the
poor quality of the data taken on 2011~July~16, arising from seeing in the range
of 2.5\arcsec compared to 0.9\arcsec on the other day.

\begin{figure*}
\centering
\includegraphics[width=17cm]{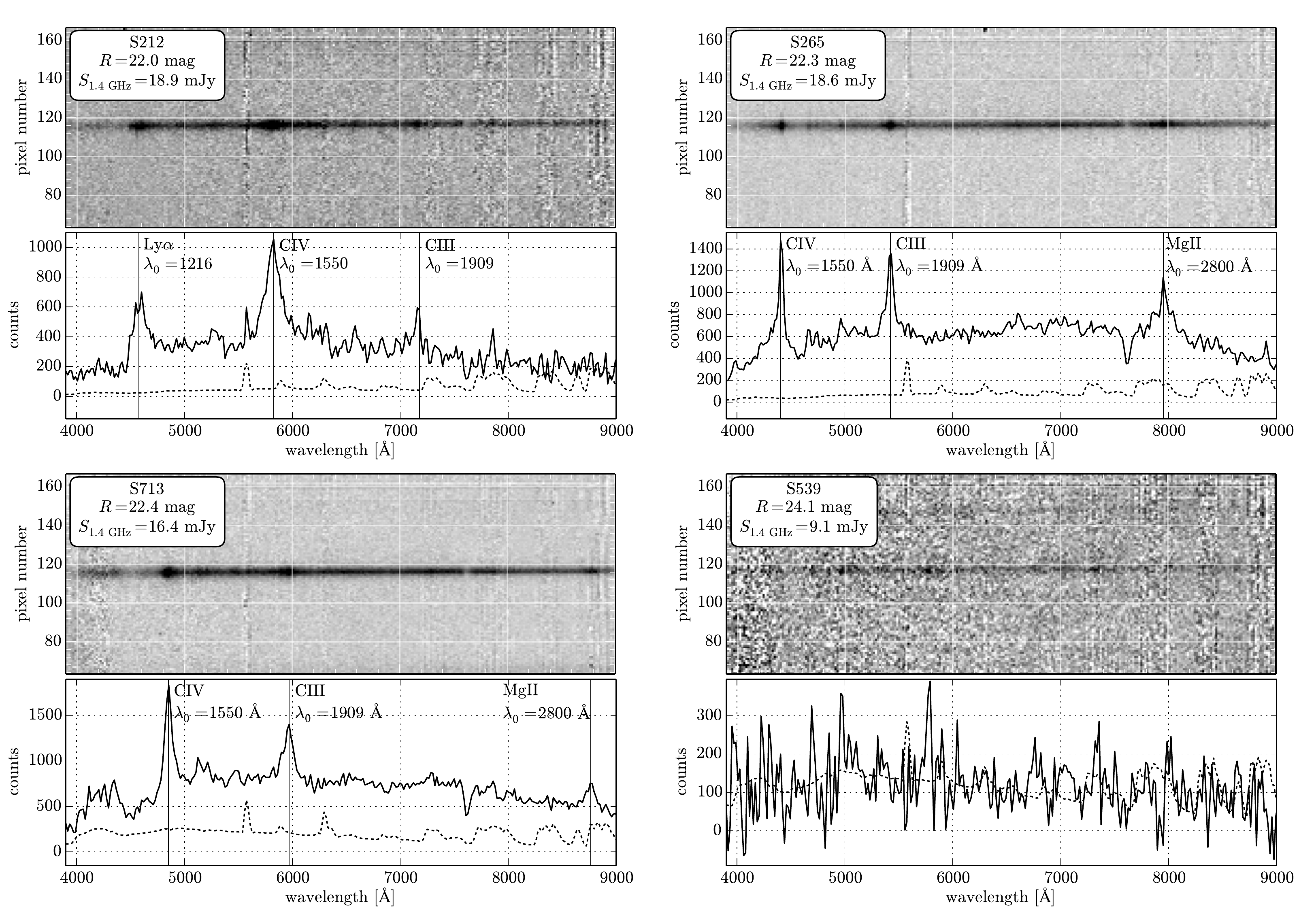}
\caption{Spectra of IFRS S212~(upper left), S265~(upper right),
  S713~(lower left), and S539~(lower right). For each IFRS we show the
  two-dimensional spectrum~(upper plot) and the extracted
  one-dimensional spectrum~(lower plot). We list the ID, the $R$~band
  Vega magnitude, and the 1.4~GHz radio flux density of each
  IFRS. In the one-dimensional spectrum, the solid line represents the
  spectrum of the IFRS while the dotted line shows the sky background
  in arbitrary units. Additionally, the position of emission lines are
  marked by vertical lines for a redshift of $z=2.76$ (S212), $z=1.84$
  (S265), and $z=2.13$ (S713).}
\label{figure:all_spectra}
\end{figure*}

The resulting one- and two-dimensional spectra of IFRS~S212, S265,
S713, and S539 are shown in Fig.~\ref{figure:all_spectra}.


\section{Redshifts and intrinsic properties of IFRS}

\label{redshifts_and_intrinsic_properties_of_IFRS}

Using the one-dimensional spectra, we measured redshifts -- where
applicable -- from the \ion{Mg}{ii} line, which is generally
considered to be the most reliable high ionisation line. Where
  \ion{Mg}{ii} was not available, we used \ion{C}{iv} to derive the
  redshift, although \ion{C}{iv} is known to be different from the
  galaxy's redshift \citep{Richards2011}. However, the effect is below
  one percent and we therefore do not observe this effect in our
  low-resolution spectra. The Ly$\alpha$ emission line is again less
  suitable for redshift determinations because of uncertainties due to
  self-absorption.\par

Since the calibration uncertainties in the spectra are negligible compared to
the uncertainties in the determination of the line position, we determined the
redshift uncertainties only from the error in the line position which was
obtained from a Gaussian fit to the emission line. We note that all redshifts
measured from all available emission lines listed in
Table~\ref{table:redshifts_widths} are within the uncertainty of our best
redshift measured from \ion{Mg}{ii} or \ion{C}{iv}.\par

Using the obtained redshifts, we calculated K-corrected radio luminosities,
assuming a power law~$S\sim \nu^\alpha$ with a radio spectral index of $-1.4$
which is the median spectral index found by \citet{Middelberg2011} for a sample
of 17~IFRS from the ATLAS sample. We also used the flattest~$(-0.7)$ and
steepest~$(-2.4)$ spectral index from \citeauthor{Middelberg2011} to constrain
the expected radio luminosity range of the observed IFRS. Hereafter, we state
these numbers in brackets. In the following, we describe the individual spectra
of our four observed IFRS, S212, S265, S713, and S539, respectively.

\subsection{S212}

Three broad emission lines with full width at half maximum~(FWHM) between
78~$\AA$ and 169~$\AA$ are visible in the spectrum of S212 which we identify as
Ly$\alpha$, \ion{C}{iv}, and \ion{C}{iii]}, at a redshift~$z=2.76\pm 0.05$ (see
Fig.~\ref{figure:all_spectra}, upper left). Furthermore, a less distinct
emission line can be associated with \ion{Si}{iv} at around 5260~$\AA$. The
broad emission lines of a few thousand km~s$^{-1}$ suggest the presence of an
AGN in S212. Using the determined redshift and the measured radio flux density
listed in Table~\ref{table:sample_IFRS}, S212 has a 1.4~GHz luminosity of
$2.0\times 10^{27}$~W~Hz$^{-1}$ $(8.8\times 10^{26}-7.6\times
10^{27}$~W~Hz$^{-1})$.

\subsection{S265}

We found three broad emission lines with FWHM between 57~$\AA$ and 223~$\AA$ in
the spectrum of S265 in Fig.~\ref{figure:all_spectra} (upper right) which we
identify as \ion{C}{iv}, \ion{C}{iii]}, and \ion{Mg}{ii}, at a
redshift~$z=1.84\pm 0.03$. These broad emission lines with line widths in the
range of 4000~km~s$^{-1}$ to 8000~km~s$^{-1}$ clearly suggest the presence of an
AGN in S265. We find a 1.4~GHz luminosity of $6.7\times 10^{26}$~W~Hz$^{-1}$
$(3.2\times 10^{26}-1.2\times 10^{27}$~W~Hz$^{-1}$).

\begin{table}

\caption{Spectroscopic information of the four IFRS observed with
  FORS2.  Listed are the IFRS ID, the spectrosopic redshift, and
  the emission line identified in the spectrum with the associated
  line width. We obtain the final redshift of each IFRS from the
  \ion{Mg}{ii} emission line and from \ion{C}{iv} where \ion{Mg}{ii} was not
  available.}
\label{table:redshifts_widths}      
\centering                          
\begin{tabular}{c c c c}        
\hline\hline                 
 IFRS & $z$ & Line & FWHM \\   
 ID &  & & [km~s$^{-1}$] \\
\hline                        
   S212 & $2.76\pm 0.05$ & \ion{C}{iv}  & $8700\pm 600$\\
        & $2.75\pm 0.02$ & \ion{C}{iii]} & $3300\pm 700$\\
   S265 & $1.85\pm 0.02$ & \ion{C}{iv}  & $3900\pm 500$ \\
        & $1.84\pm 0.02$ & \ion{C}{iii]} & $4200\pm 300$\\
        & $1.84\pm 0.03$ & \ion{Mg}{ii} & $8400\pm 900$\\
   S713 & $2.13\pm 0.03$ & \ion{C}{iv}  & $5400\pm 700$ \\
        & $2.12\pm 0.03$ & \ion{C}{iii]} & $5900\pm 400$\\
        & $2.13\pm 0.03$ & \ion{Mg}{ii} & $4600 \pm 1000$\\
   S539 & -- & -- & -- \\
\hline                                   
\end{tabular}
\end{table}

\subsection{S713}
Three broad emission lines are visible in the spectrum of S713, with FWHMs
between 884~$\AA$ and 134~$\AA$, corresponding to line widths about
5000~$\mathrm{km}~\mathrm{s}^{-1}$ and suggesting the presence of an AGN. The
lines are associated with \ion{C}{iv}, \ion{C}{iii]}, and \ion{Mg}{ii} at a
redshift~$z=2.13\pm 0.03$ (see Fig.~\ref{figure:all_spectra}, lower left). We
find a 1.4~GHz luminosity of $8.7\times 10^{26}$~W~Hz$^{-1}$ $(3.9\times
10^{26}-2.7\times 10^{27}$~W~Hz$^{-1})$.

\subsection{S539}
The source~S539 is the optically faintest IFRS in this observing
programme. Unfortunately, it was only observed with 44~min
on-source time, although 3~h had been requested. Therefore, the
resulting spectrum does not provide the quality of the other spectra
and we cannot use it to measure a redshift and line widths (see
Fig.~\ref{figure:all_spectra}, lower right). Despite the short
integration time, emission features are visible, although with
poor signal-to-noise. We tentatively interpret the emission
feature at around 4970~$\AA$ as Ly$\alpha$, indicated by the
related break towards lower wavelengths, and a second emission feature
at redder wavelengths as \ion{Si}{iv}. We suggest S539 to be at
redshift $z\sim 3.1$, although the low signal-to-noise ratio prevents
a reliable determination.

\section{Discussion}

\label{discussion}

We have obtained for the first time spectroscopic redshifts for three out of
four IFRS in the ATLAS fields selected on the basis of their existing optical
counterparts. All spectra provide solid determinations of redshifts between 1.8
and 2.8, providing strong evidence that these sources are located at high
redshifts. The fourth IFRS has a low signal-to-noise ratio spectrum that
indicates a redshift of 3.1 but needs additional confirmation. These redshifts
are in agreement with the conclusions of \citet{GarnAlexander2008},
\citet{Huynh2010}, \citet{Norris2011}, and \citet{Zinn2011} who suggested that
IFRS are located at redshifts above~2, mainly from SED modelling. As mentioned
in Sect.~\ref{sample_and_observations}, we are aware that the selection of the
optically brightest IFRS might bias our sample towards lower redshifts. This
seems to be in agreement with the measured redshifts, which are in the lower
part of the expected redshift range of IFRS. Recently, \citet{Collier2014} presented
19~IFRS with spectroscopic redshifts from archival data. Their sample was
extracted from a shallow all-sky survey, in contrast to the IFRS analysed in
this paper which were found in the deep ATLAS fields. Therefore, the IFRS found
by \citeauthor{Collier2014} were radio brighter and slightly IR~brighter than
the ones presented in this work. Nevertheless, \citeauthor{Collier2014} found
redshifts in the range $2<z<3$, in agreement with those presented here.\par

All spectra shown in Fig.~\ref{figure:all_spectra} are broad-line
quasar spectra, characterised by high-ionisation emission lines with
high equivalent widths. This finding agrees with former suggestions by
\citet{GarnAlexander2008}, \citet{Huynh2010}, \citet{Norris2011}, and
\citet{Zinn2011} that IFRS contain AGN.\par

\subsection{Similarity between IFRS and HzRGs}
\label{similarityIFRS_HzRGs}

In Sect.~\ref{redshifts_and_intrinsic_properties_of_IFRS}, we derived
1.4~GHz radio luminosities between $6.7\times 10^{26}$~W~Hz$^{-1}$ and
$2.0\times 10^{27}$~W~Hz$^{-1}$ $(3.2\times 10^{26}-7.6\times
  10^{27}$~W~Hz$^{-1}$) for the IFRS investigated in this work. The
classical separation between \citet{FanaroffRiley1974} types~1 and 2 is
$4.8\times 10^{25}$~W~Hz$^{-1}$ ($1.7\times 10^{25}$~W~Hz$^{-1}$) at
1.4~GHz, using a steep (ultrasteep) radio spectral
index~$\alpha=-0.8~(-1.3)$ for the conversion from the 178~MHz
luminosity given by \citeauthor{FanaroffRiley1974}. This clearly
classifies IFRS as Fanaroff~\&~Riley type~2.\par

\citet{Seymour2007} defined an HzRG as a radio galaxy with $z>1$ and a
3~GHz luminosity above $10^{26}$~W~Hz$^{-1}$, corresponding to
$1.8\times 10^{26}$~W~Hz$^{-1}$ ($2.7\times 10^{26}$~W~Hz$^{-1}$) at
1.4~GHz. We find the IFRS in agreement with the radio luminosity range
of HzRGs for the entire range of spectral indices of IFRS
  found by \citet{Middelberg2011}, supporting the hypothesis of
\citet{Norris2011} that IFRS might be siblings of HzRGs. We note that
most IFRS are radio-brighter and possibly more radio-luminous than
those analysed in this work.\par

\begin{figure}
\centering
         \includegraphics[width=\hsize]{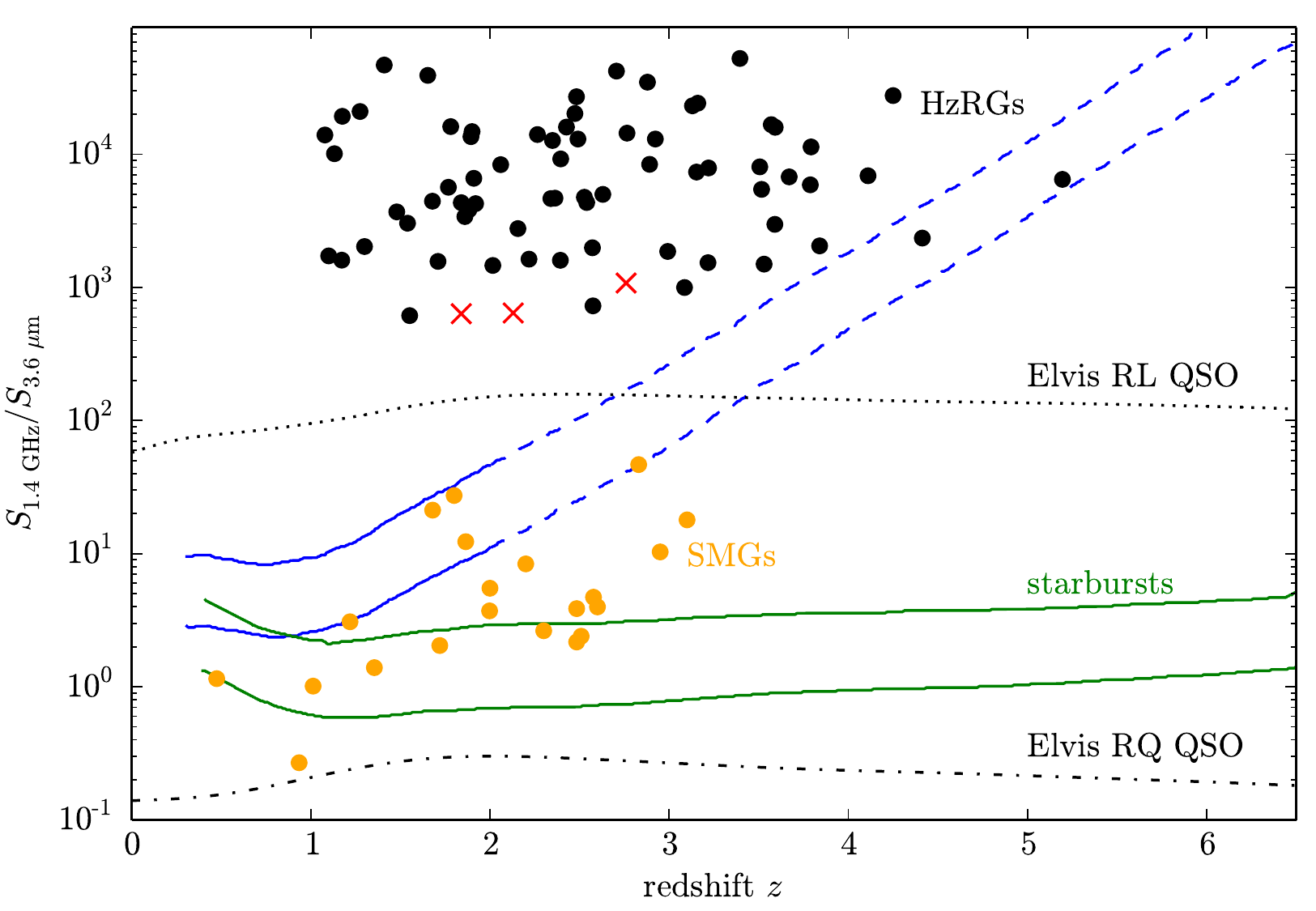}
    \caption{Ratio of 1.4~GHz and 3.6~$\mu$m flux densities for IFRS
      and several other classes of object as a function of redshift,
      adapted from \citet{Norris2011}. It shows that the IFRS analysed
      in this paper (red crosses) are more similar to HzRGs
      (black dots, \citealp{Seymour2007}) than to other types of
      galaxies frequently found at high redshifts. The solid
      lines indicate the expected loci of luminous and ultra-luminous
      infrared galaxies~(ULIRGs), using the templates from
      \citet{Rieke2009}. The dotted and dot-dashed lines indicate the
      loci of a classical radio-loud and radio-quiet QSO,
      respectively, from \citet{Elvis1994}. The location of classical
      submillimetre galaxies is indicated by the orange
      dots. We note that dust extinction could cause any of the
      calculated tracks to rise steeply at high redshift, where the
      observed $3.6~\mu$m emission is generated in visible wavelengths
      in the galaxy rest frame. This is illustrated by the dashed
      lines which show the effect of adding ${\rm A_v} = 8^m$ of
      extinction to the two starburst tracks. However, the radio
      emission from these galaxies would then be undetectable
      at $z>2$ with current sensitivity.}
    \label{figure:ratio-redshift}
\end{figure}

Infrared-faint radio sources are characterised by their extremely high
radio-to-IR flux density ratios typically in the range of several hundreds to a
few thousands. The redshifts determined in
Sect.~\ref{redshifts_and_intrinsic_properties_of_IFRS} enable us to put these
three IFRS in the plot showing the radio-to-IR flux density versus redshift
(Fig.~\ref{figure:ratio-redshift}). This plot clearly separates IFRS from other
classes of object which are typically found at high redshifts and indicates a
potential connection between IFRS and HzRGs.\par

Based on this finding, \citet{Norris2011} suggested that IFRS might follow a
relation between redshift and 3.6~$\mu$m flux density, similar to the correlation found
for the sample of HzRGs by \citet{Seymour2007}. We can now test this hypothesis.

\begin{figure}
\centering
         \includegraphics[width=\hsize]{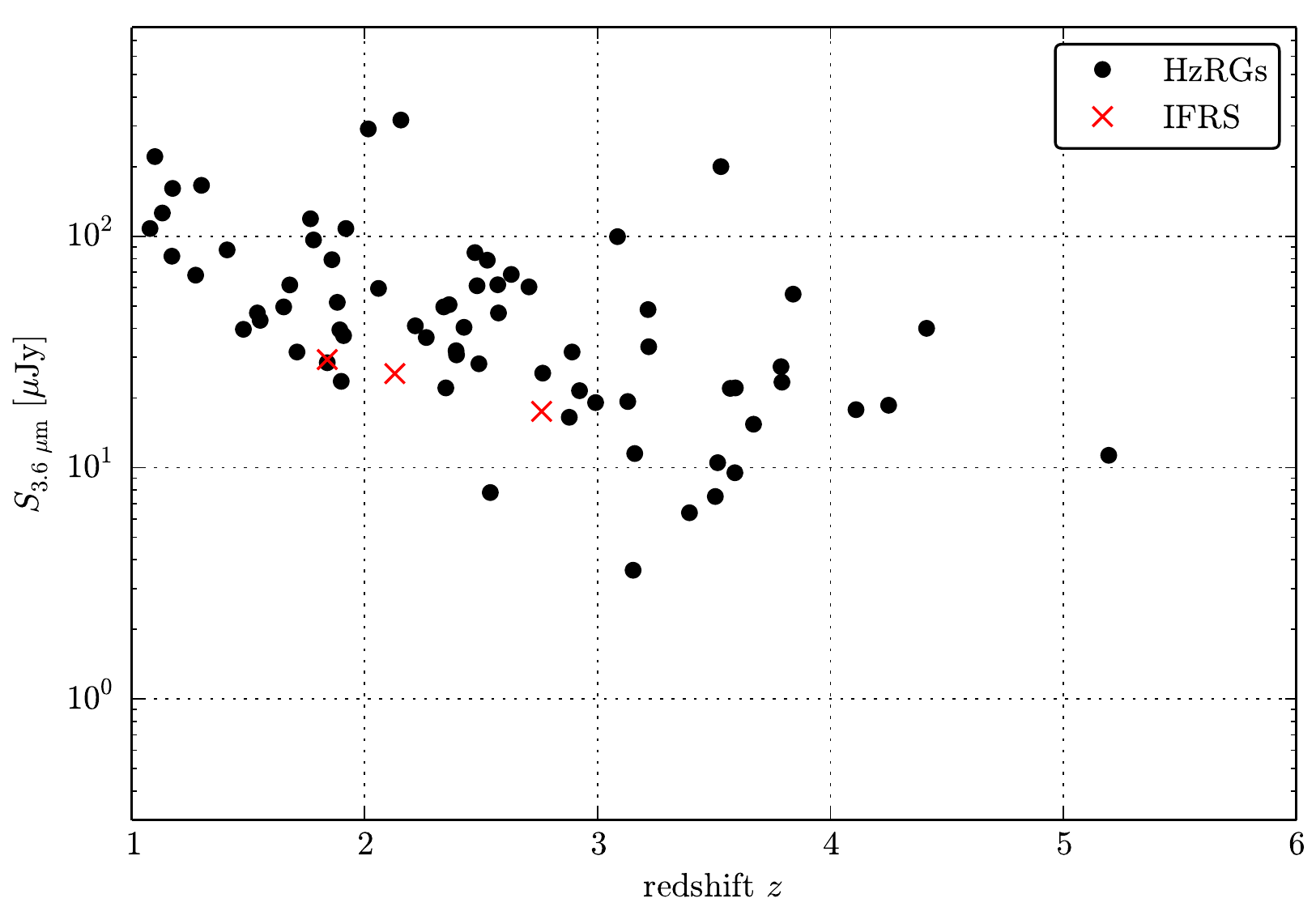}
   \caption{3.6~$\mu$m IR flux density versus redshift, adapted from
   \citet{Norris2011}. Shown is the sample of HzRGs from
   \citet{Seymour2007} as black dots and the three IFRS whose redshifts are
   presented in this work as red crosses. The IFRS are located in
   the same parameter range as the HzRGs and seem to follow their
   $S_{3.6~\mu\mathrm{m}}-z$ relation. It appears that IFRS define the
   lower bound of HzRGs.}
      \label{figure:redshift-IRflux}
\end{figure}

Figure~\ref{figure:redshift-IRflux} shows the 3.6~$\mu$m IR flux density versus
redshift for the three IFRS whose redshifts we determined in
Sect.~\ref{redshifts_and_intrinsic_properties_of_IFRS} and for the HzRGs from
\citet{Seymour2007}. Our IFRS clearly fall in the parameter space of HzRGs,
supporting the hypothesis of \citeauthor{Norris2011} in the tested redshift
range between 1.8 and 2.8. This result provides evidence that the correlation
works to redshifts of 2 or 3. It seems that IFRS define the lower bound of HzRGs
in this plot.

\subsection{Redshift-based SED modelling}

The availability of redshifts for IFRS now allows us to
perform a more accurate SED
modelling. \citet{GarnAlexander2008}, \citet{Huynh2010}, and
\citet{Zinn2011} modelled SEDs to constrain the redshift of IFRS.
\begin{figure}
\centering
         \includegraphics[width=\hsize]{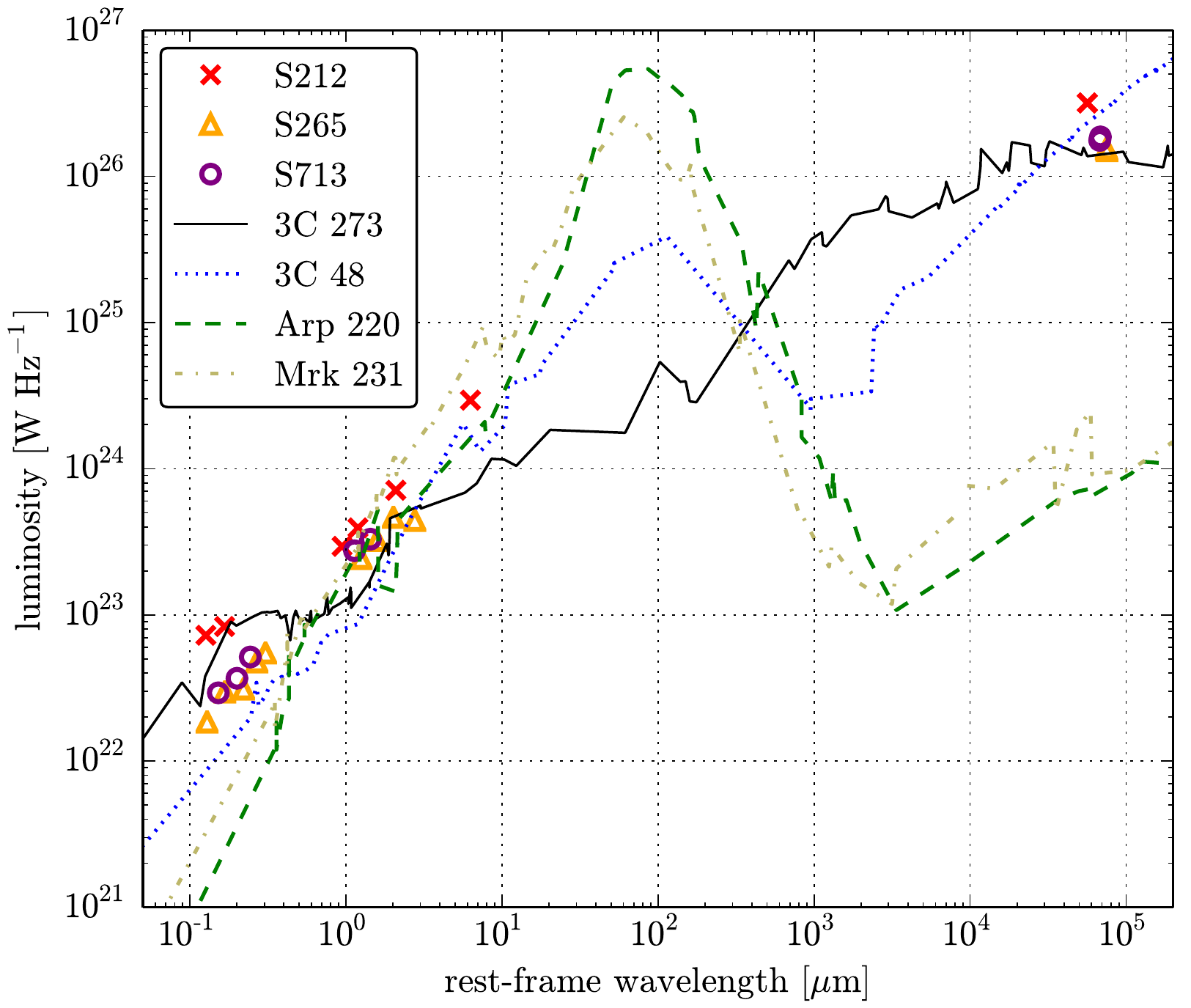}
   \caption{Rest-frame SED modelling for the three IFRS whose
     redshifts were determined in this work. All templates are
     shifted to the rest-frame and scaled in luminosity to match the
     photometric data points shown by red crosses~(S212), orange
     triangles~(S265), and purple circles~(S713). While the SED
     templates of star forming galaxy Arp\,220~(green dashed line) and
     Seyfert galaxy Mrk\,231~(olive dashed-dotted line) clearly
     disagree with the available photometric data, the templates of
     the radio-loud quasar 3C\,273~(black solid line) and the compact
     steep spectrum~(CSS) source 3C\,48~(blue dotted line) fulfill the
     requirements.}
      \label{figure:SED_fitting}
\end{figure}

Here, in contrast, we can use the redshift as an anchor and use it to
test different SED templates against available photometric
data. The method of our SED modelling is similar to the
  approaches by \citet{Huynh2010} and \citet{Zinn2011}.\par

We built the SED templates using photometric data and redshifts
from the NASA/IPAC Extragalactic Database~(NED), connecting the
datapoints by lines and smoothing the template. We used a
variety of SED templates, including starburst, radio and dwarf
galaxies, and quasars. Furthermore, we took all available
photometric data of the IFRS from \citet{Norris2006}. These data
consist of optical ($G$ and $R$~bands), IR~(3.6~$\mu$m and 4.5~$\mu$m),
and radio~(1.4~GHz) flux densities for all three IFRS, while S212 and
S265 provide more data points in the optical and IR range.\par

To model the SEDs of IFRS, we shifted the template SEDs to the
rest-frame and scaled them in luminosity to match the observed
3.6\,$\mu$m flux density of the IFRS. Extinction was added in the
rest-frame optical and near-IR, following a \citet{Calzetti2000} reddening law,
where required by the photometric data. Furthermore, all available photometric
datapoints of the IFRS from NED were also shifted to the IFRS
rest-frames and used to test the template SEDs for compatibility with these
available constraints.\par

Figure~\ref{figure:SED_fitting} shows the SED modelling for the three IFRS. We
find that S265 and S713 have very similar SEDs; Infrared-faint radio source S212
is also similar, except that it is a factor of~2 or 3 brighter than the other
two at all wavelengths. These SEDs are consistent with either the 3C\,273 or the
3C\,48 templates. In our modelling, we scaled down these two templates in
luminosity by wavelength-independent factors of 17 and 10, respectively, to
match the observed 3.6\,$\mu$m flux densities. In the case of the templates of
3C\,273 and 3C\,48, no adding of extinction was needed to match the available
flux densities.\par

Except for 3C\,273 and 3C\,48, none of the templates applied matches the SEDs of
the IFRS, even with reddening applied. This result is in agreement with the
finding that IFRS are high-redshift radio-loud AGN as confirmed by the spectra
shown in Fig.~\ref{figure:all_spectra}. Our SED modelling is inconsistent with
the alternative interpretation of IFRS mentioned by \citet{Norris2011},
explaining IFRS as AGN that undergo heavy dust extinction.\par

\subsection{Are IR-fainter IFRS at higher redshifts? Cosmological
relevance of IFRS}

In Sect.~\ref{similarityIFRS_HzRGs}, we showed that our results provide evidence
for the similarity between IFRS and HzRGs. In particular, we found our IFRS to
be in agreement with the redshift--$3.6\,\mu$m flux density relation determined
for HzRGs by \citet{Seymour2007} in the redshift range probed by our sample.
However, our sample was essentially determined by observational constraints,
since fainter sources would have required unreasonably long integration times on
the largest facilities.\par
  
If we assume that IFRS follow the relation for at least somewhat higher
redshifts, then most IFRS in the ATLAS fields would be at even higher
redshifts, potentially reaching 5 or 6, since the IFRS analysed in this paper
are the optically and IR-brightest IFRS in the ATLAS fields (see
Sect.~\ref{sample_and_observations}). Whilst measuring such redshifts in the
optical regime requires prohibitive amounts of observing time, the detection of
molecular lines such as CO at radio frequencies might be much more efficient.
The Atacama Large Millimeter Array~(ALMA) is therefore the instrument of choice
with which to test this hypothesis of an extension of the redshift--$3.6\,\mu$m
flux density relation towards higher redshifts.\par

We have shown that IFRS have similar properties to HzRGs, but they have a
significantly higher sky density of a few per square degree. If the IR-fainter
IFRS are indeed at even higher redshifts, the number of AGN in the early
universe would be larger, resulting in even greater problems with structure
formation and the growth of SMBHs shortly after the Big Bang (see
\citealp{Haiman2013} for a recent review). This had already been emphasised by
\citet{Zinn2011} under the assumption that IFRS are AGN~driven and are located
at redshifts of 3. Since our evidence supports their premises, we conclude that
IFRS are likely to be cosmologically relevant.\par


\section{Conclusions}

\label{conclusions}

We present the first spectroscopic data of four IFRS in the ATLAS
fields from the VLT~FORS2 and determine the properties of
IFRS.

\begin{itemize}

  \item We determined the first redshifts of ATLAS-IFRS and find three
    IFRS at $z=1.84$, 2.13, and 2.76, providing evidence of
    the suggested high redshifts of IFRS.

  \item Broad emission lines with line widths between 3300\,km\,s$^{-1}$ and
    8700\,km\,s$^{-1}$ found in all spectra substantiate the claim
    that IFRS contain AGN.

  \item Using the redshifts measured in this work, we present the
    first SED modelling of IFRS based on redshifts and find the
    template SEDs of radio-loud quasars to agree with that of IFRS.

  \item IFRS have derived radio luminosities similar to those of
    HzRGs, providing further evidence of the similarity of IFRS and
    HzRGs.

  \item We test the hypothesis that IFRS follow the same
    correlation between 3.6~$\mu$m flux density and redshift as
    HzRGs. Our findings support this hypothesis in the tested redshift
    range and increase the likelihood that IR-fainter IFRS are at even
    higher redshifts, potentially reaching 5 or 6. Considering their
    sky density of a few per square degree, IFRS would significantly
    increase the number of AGN in the early universe, leading to even
    more problems with structure formation and the growth of SMBHs
    shortly after the Big Bang.
\end{itemize}

\begin{acknowledgements}
      We thank Peter-Christian Zinn for providing the basics of the
      code for the SED modelling. AH acknowledges funding from \textit{
      Bundesministerium f\"ur Wirtschaft und Technologie} under the label
     50\,OR\,1202. The author is responsible for the content of this
     publication. Based on observations made with ESO Telescopes at the La Silla
     Paranal Observatory under programme ID~087.B-0813(A). IRAF is distributed
     by the National Optical Astronomy Observatory, which is operated by the
     Association of Universities for Research in Astronomy~(AURA) under
     cooperative agreement with the National Science Foundation. This research
     has made use of the NASA/IPAC Extragalactic Database~(NED) which is
     operated by the Jet Propulsion Laboratory, California Institute of
     Technology, under contract with the National Aeronautics and Space Administration.
\end{acknowledgements}


\bibliographystyle{aa} 
\bibliography{references} 

\end{document}